\documentclass[osajnl,preprint,showpacs,superscriptaddress,12pt]{revtex4-1} %% use 12pt for preprint option
\usepackage{amsmath,amssymb,graphicx}
\usepackage[figuresleft]{rotating}
\begin{document}

\title{Curvature Wavefront Sensing for the Large Synoptic Survey
  Telescope}

\author{Bo Xin}\email{Corresponding author:bxin@lsst.org}
\author{Chuck Claver}
\author{Ming Liang}
\author{Srinivasan Chandrasekharan}
\author{George Angeli}
\affiliation{Large Synoptic Survey Telescope, 933 N Cherry Ave.,
Tucson, AZ 85719, USA}

\author{Ian Shipsey}
\affiliation{Department of Physics, University of Oxford, Oxford, UK}

\begin{abstract}
The Large Synoptic Survey Telescope (LSST) will use an active optics
system (AOS) to maintain alignment and surface figure on its
three large mirrors. Corrective actions fed to the LSST AOS are
determined from information derived from 4 curvature wavefront sensors located at the
corners of the focal plane. Each wavefront sensor is a split detector such that the halves
are 1mm on either side of focus. In this paper we describe the
extensions to published curvature wavefront sensing algorithms needed
to address challenges presented by the LSST, namely the large central
obscuration, the fast $f$/1.23 beam, off-axis pupil distortions, and vignetting
at the sensor locations.   We also describe corrections needed for the
split sensors and the effects from the angular separation of different
stars providing the intra- and extra-focal images. Lastly, we present simulations
that demonstrate convergence, linearity,
and negligible noise when compared to atmospheric effects when the
algorithm extensions are applied to the LSST optical system.  The
algorithm extensions reported here are generic and can easily be
adapted to other wide-field optical systems including similar
telescopes with large central obscuration and off-axis curvature sensing.
\end{abstract}


\maketitle %% required

\section{Introduction}

\label{sec:intro}  % \label{} allows reference to this section
The Large Synoptic Survey Telescope (LSST) is a new facility now under
construction that will survey $\sim$20000 square degrees of the
southern sky through 6 spectral filters ($ugrizy$) multiple times over
a 10-year period~\cite{Ivezic08}.  The conduct of the LSST survey is 
defined by a 2-exposure ``visit'' lasting $\sim$39 seconds having
 the following pattern:  15s integration plus 1s shutter open/close transitions (16s 
 elapsed time), 2s focal plane array (FPA) readout, 16s exposure, 5s 
telescope repointing and FPA readout (the last 2s readout is 
concurrent with the telescope repointing).  With this cadence the LSST
will observe $\sim$800 visits per 8 hour night, producing more than 1600 
images and $\sim$15 Tbytes per night of data and cover the entire visible sky
every 3-4 nights.  Construction of the LSST includes an 8-meter class 3-mirror
telescope, a 3.2 billion pixel camera, and an extensive computing
system for data analysis and archiving.  Commissioning of the LSST is
expected to begin in late 2019 and full surveying operations at the end of 2022. 

\begin{figure*}
  \begin{center}
  \begin{tabular}{c}
\includegraphics[width=150mm,height=80mm]{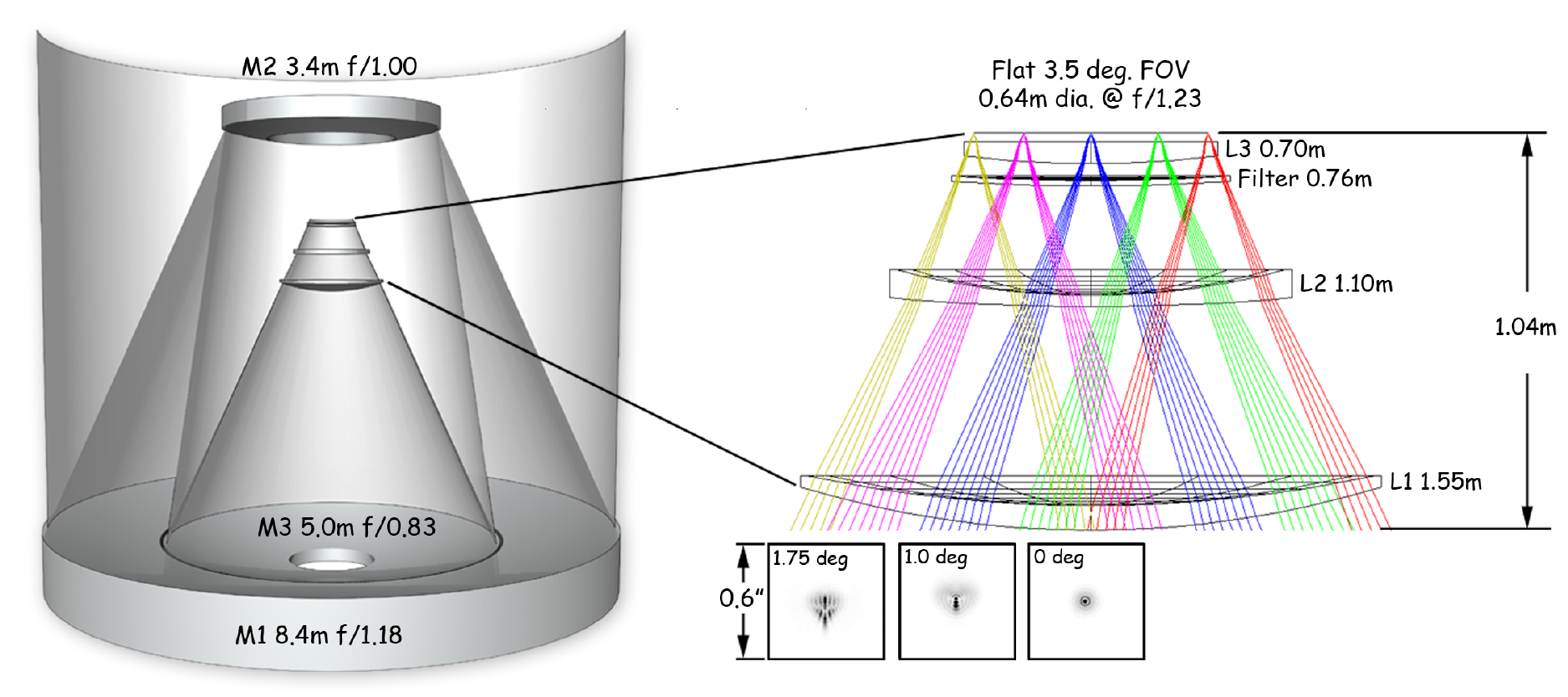}
 \end{tabular}
  \end{center}
  \caption 
 { \label{fig:optics} 
The 3-mirror telescope optical configuration for the LSST (left)
showing the placement of the tertiary mirror within the primary
mirror,  allowing the two to be fabricated from a single substrate.
The camera optics (right) use three fused-silica lenses and a meniscus
filter.  The r-band image quality (lower right) of the LSST optical
system is shown for 0, 1.0 and 1.75 degrees from the optical axis.
}
  \end{figure*}

The optical system (Figure~\ref{fig:optics}, left) of the LSST is
based on a modified Paul-Baker 3-mirror
telescope design having an 8.4m primary, 3.4m secondary, and 5.0m
tertiary feeding a three-element refractive camera system producing a
flat 3.5 degree field-of-view with an effective clear aperture of
6.5m.  The mean intrinsic imaging performance of the optical design
across the optical spectrum from 320nm to 1150nm is better than
0.1 arcsecond full width at half maximum (FWHM) over the full field of
view. The optical design has allowed the primary and tertiary mirrors to be fabricated out of a
single substrate.  Fabrication of the primary-tertiary (M1M3) mirror pair has
recently been completed by the Steward Observatory Mirror Lab. 

The overall system image quality budget for the LSST is 0.4
arcsecond FWHM and is allocated between the telescope (0.25") and camera (0.3")
subsystems. With this image quality budget the LSST's delivered image
quality is dominated by the atmospheric seeing at its site on Cerro
Pachon in Chile.  For the telescope subsystem, the bulk of its budget
allocation is taken up by  residual mirror figure errors and
thermal and gravity-induced misalignments of the camera and secondary mirror
systems.  In order to maintain this image quality budget the LSST telescope 
subsystem utilizes an active optics system (AOS) controlling 50 degrees
of freedom (DoF) consisting of 20 bending modes each on the actively
supported M1M3 and M2 systems and 5 DoF of rigid body position for the
camera and M2.  By contrast, in the camera subsystem much of its budget
allocation is taken up by internal charge diffusion of the  
100$\mu$m-thick CCDs and overall flatness of the FPA. 

Most modern astronomical telescopes use some form of AOS with 
near-realtime optical wavefront sensing feedback.  The most common method 
for estimating the wavefront in these systems is with a Shack-Hartmann 
wavefront sensor (SHWFS), e.g. the Magellan telescopes~\cite{magellan}
and the VLT~\cite{vlt}.  Wide 
field systems with their fast optical beams make using Shack-Hartmann
wavefront sensing problematic for two reasons: 1) To avoid the vignetting
effects a pickoff mirror would 
cause in a fast converging optical beam, the SHWFS would have to be
placed near the focal surface. This means these mechanisms would
have to be small and within the vacuum cryostat of the imager.
2) Some form of articulation
is needed to place the reference star within the SHWFS limited field
of view.  

Currently operating wide field survey telescopes all use wavefront
sensing based on image analysis from area detectors. The VISTA telescope, an
infrared survey telescope, utilizes a pair of curvature sensors and 
forward modeling of low order aberrations to provide optical
feedback~\cite{vista1,vista2}.  Similarly, the Dark Energy
Camera employs 8 2k$\times$2k sensors around the focal plane both in and out
of focus to forward model the resulting donut
images~\cite{roodman14}.  
Curvature wavefront sensing is also in use on several operating telescopes   
including the 3.5m WIYN~\cite{wiyn95} and 
4m Mayall~\cite{mayall00} 
telescopes at Kitt Peak National  Observatory.  These
telescopes use curvature wavefront sensing to  
build their AOS look-up tables and to periodically set the  
operational zero points for the AOS.  
In each of these systems the number of DoF controlled is
limited to focus, misalignment of the instrument or secondary mirror
(coma), and one or two forms of primary mirror bending (astigmatism and/or trefoil).

Because LSST uses two actively supported mirror systems and two
positioning hexapods, the number of controlled DoF
is significantly greater than typically seen in currently operating telescopes.
Controlling these DoF requires the ability to estimate higher order properties of
the aberrated wavefront,  Zernike coefficients Z4 -- Z22 
in Noll/Mahajan's definition~\cite{standardZ,annularZ}. 
Further,  the time constraints imposed by the LSST's rapid cadence requires that
the process of measuring the optical wavefront and maintaining alignment and 
figure control be highly automated and reliable.  For these reasons we
have adopted curvature wavefront sensing to provide the optical feedback
needed to control the LSST's AOS.  

\begin{figure*}
  \begin{center}
  \begin{tabular}{c}
\includegraphics[width=150mm,height=80mm]{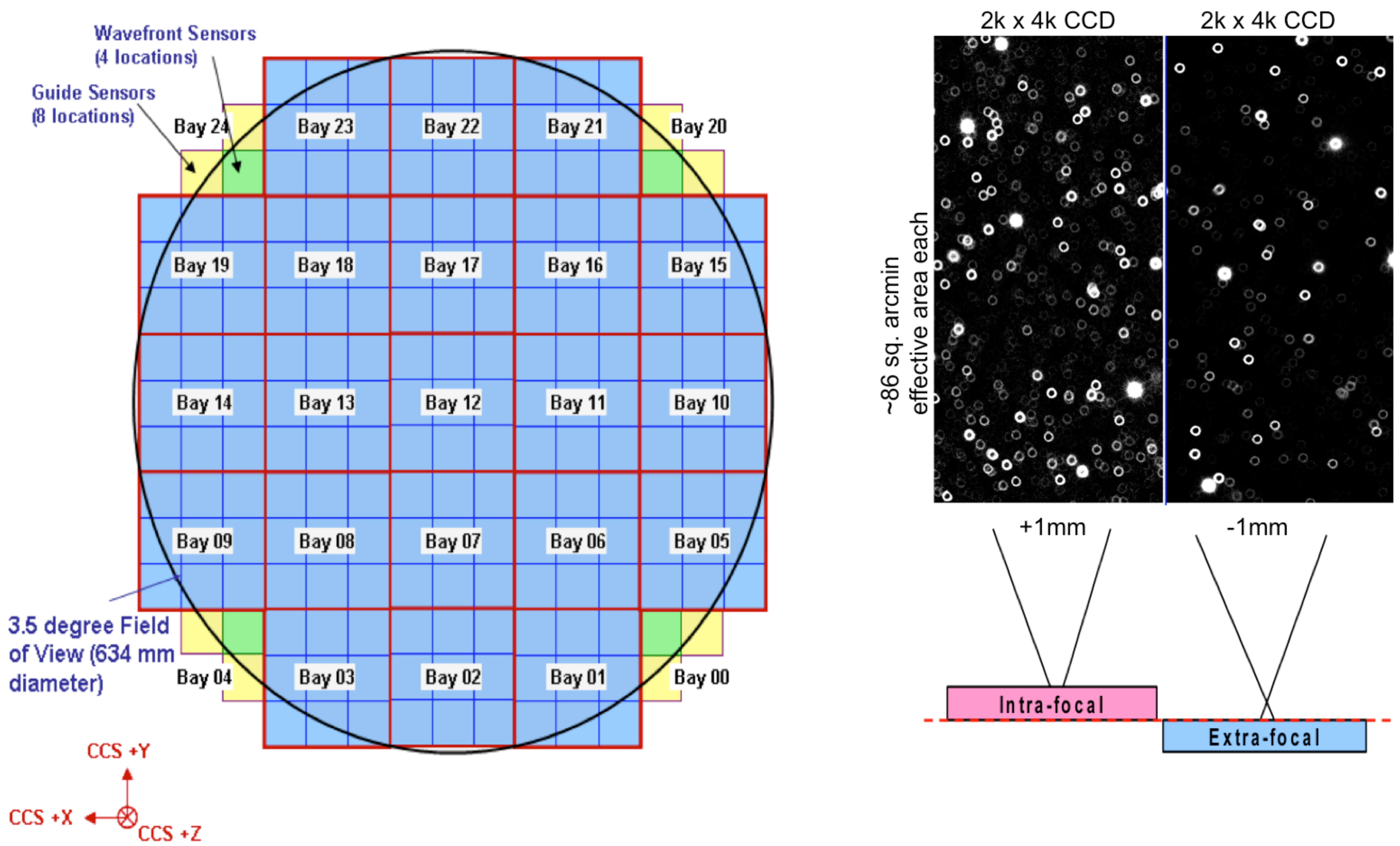}
 \end{tabular}
  \end{center}
  \caption 
 { \label{fig:sensorlayout} 
The focal plane configuration for the LSST (left panel) showing 
sensors used for science (blue), guiding (yellow), and wavefront sensing 
(green).  Each raft is mounted into a preadjusted grid ``bay''.
The wavefront sensors are divided into 
two halves of intra- and extra-focal sensors (right panel). 
The strategy in making use of multiple stars on each half-chip to
obtain a wavefront measurement at each corner is currently under
investigation and not discussed in this paper. 
}
  \end{figure*}

The layout of the LSST focal
plane (Figure~\ref{fig:sensorlayout}, left) was developed in part to
accommodate the LSST's wavefront sensing requirements. The science
sensors (blue) are arranged in 21 modular ``rafts'', each containing 9
CCDs.   At four
locations are ``corner rafts'' consisting of two guide sensors
(yellow) and one wavefront sensor (green) each.  By measuring the
field-dependent Z4 --
Z22 at 4 locations we have 76 variables to control 50 DoF.

Curvature wavefront sensing offers some advantages over other
wavefront sensing methods for wide field  survey telescopes.  By
relying on equally defocused intra- and extra-focal images, curvature
sensing can use area sensors with relatively large fields-of-view.  
This allows significant flexibility in selecting reference sources to 
use for the wavefront measurement when the source scenery is
constantly changing from one visit to the next with the LSST cadence. 

Typically this is achieved
through using a beam splitter and delay line or by physically moving
the detector.  Both of these approaches are problematic for the LSST
with its fast $f$-number ($f$/1.23) and crowded focal plane. 
Our design for LSST therefore splits the available wavefront sensor area in two
halves (see Figure~\ref{fig:sensorlayout}, right), with one 1mm in
front of nominal focus and the other 1mm behind.  Each half utilizes a
2k $\times$ 4k CCD with 10$\mu$m square pixels providing a 7$\times$14
arcminute field of view.  With this field of view, the likelihood of acquiring suitable
reference stars is near unity even in the low density galactic pole
regions~\cite{manuel2010}.  Therefore, no active acquisition of the 
reference source is required.

There are a variety of algorithms used to estimate the wavefront in
curvature wavefront sensing. However, none of them by themselves 
can work with a wide-field telescope having large central obscuration and off-axis
sensors like LSST.  Our strategy is to choose two
well-established algorithms that are known to  work for large
$f$-number, on-axis systems, implement them as our baseline, 
then extend them to work with small $f$-number and off-axis sensors.
As our two baseline algorithms we have chosen the iterative Fast Fourier 
Transform (FFT) method by Roddier and Roddier~\cite{Roddier93}, and
the series expansion technique by Gureyev and Nugent~\cite{GuNu96}.
We found both methods to be accurate and reasonably fast.
We note that our extensions to the baseline algorithms can also be used with
other curvature sensing algorithms that work for large $f$-number
and on-axis systems, in order to make them work for small $f$-number and
off-axis sensors.

In this paper, we describe the development of the curvature wavefront
sensing algorithm that measures the wavefront at the 4 corner locations. 
The paper is organized as follows.
In Section~\ref{sec:baseline} we review our baseline algorithms of
curvature wavefront sensing,
and in Section~\ref{sec:extension} we discuss the new challenges facing a wide-field and
off-axis system like LSST. We also discuss in
Section~\ref{sec:extension} the required modifications to the 
baseline algorithms in order to overcome these challenges.
Section~\ref{sec:tests} then gives simulation results of unit testing
and validations, including  
analyses of algorithmic noise and atmospheric
background.

\section{Baseline Curvature Sensing Algorithms}
\label{sec:baseline}

The concept of curvature wavefront sensing was first developed and demonstrated by 
F. Roddier~\cite{Roddier88}. 
The underlying idea is to measure the spatial intensity
distribution of a star at two positions, one on either side of
focus. 
The derivative of the local surface brightness of the defocused images
along the direction of propagation is
given by the transport of intensity equation (TIE)~\cite{Roddier93}:
\begin{equation}
\frac{\partial I}{\partial z} =  -(\nabla I \cdot \nabla W + I
\nabla^2W), 
\label{eq:tie}
\end{equation}
where $I$ is the intensity, $W$ is the wavefront error, and $z$ is the
distance 
between the conjugate planes of the defocused pupil images.

Assuming the intensity is equal to $I_0$ everywhere inside the pupil
and zero outside, we have the Neumann boundary conditions:
\begin{equation}
\nabla I  =  -I_0 \hat{n} \delta_c,
\end{equation}
where $\delta_c$ is a delta function around the pupil edge.
Therefore, the first term in Eq.~(\ref{eq:tie}) is localized at the
beam edge. Across the beam, the derivative of the local image
brightness is 
proportional to the Laplacian, or curvature, of the wavefront.

As a partial linear differential equation, the TIE can be solved with
various methods.  In the following 2 subsections we summarize the two
methods we have adopted as our baseline algorithms.

\subsection{Iterative FFT}

The iterative FFT algorithm~\cite{Roddier91} solves
the TIE by making use of the Laplacian operator
becoming a simple arithmetic operation in Fourier space.

The longitudinal derivative of the intensity can
be expressed as 
\begin{equation}
-\frac{1}{I_0}
\frac{\partial I}{\partial z} =  \nabla^2W -  \delta_c \frac{\partial
 W}{\partial \vec{n}}.
\label{eq:tie1}
\end{equation}
We define the wavefront signal $S$ using 
the longitudinal derivative normalized by $I_0$. As such, $S$ can be approximated as:
\begin{equation}
S=-\frac{1}{I_0}
\frac{\partial I}{\partial z} 
\approx  -\frac{I_1-I_2}{I_0 \cdot   2\Delta z}
\approx -\frac{1}{\Delta z} \frac{I_1-I_2}{I_1+I_2},
\label{eq:pIpz}
\end{equation}
where the focus offset in the object space (e.g. the pupil) is given
by 
\begin{equation}
\Delta z =f(f-l)/l,
\label{eq:deltaz}
\end{equation} 
with $f$ the system focal length, and $l$ the
defocus distance of the intra/extra focal image planes.

It can be shown that if we constrain $\partial W/\partial \vec{n}$ on
the edge, $\delta_c$ can be absorbed into the Laplacian.
An estimate of the Laplacian of the wavefront error can be rewritten
as
\begin{equation}
\nabla^2 W \approx S.
\end{equation}
Using the following property of the Fourier Transform (FT) (see for example page 314 of Ref.~\citenum{Gaskill}),
\begin{eqnarray}
FT({\mu,\nu})\{ \nabla^2 W(x,y)\} & = & -4\pi^2 (\mu^2+\nu^2) \nonumber\\
& & FT({\mu,\nu})\{ W(x,y)\},
\end{eqnarray}
where $\mu$ and $\nu$ are the spatial frequencies, we can solve for $W$ using the inverse Fourier Transform ($IFT$):
\begin{equation}
W\approx IFT({x,y}) \left\{ \frac{FT({\mu,\nu})\{S\}}{-4\pi^2(\mu^2+\nu^2)}
\right\}.
\label{eq:wcs}
\end{equation}

\begin{figure*}[tb]
  \begin{center}
  \begin{tabular}{c}
\includegraphics[width=160mm]{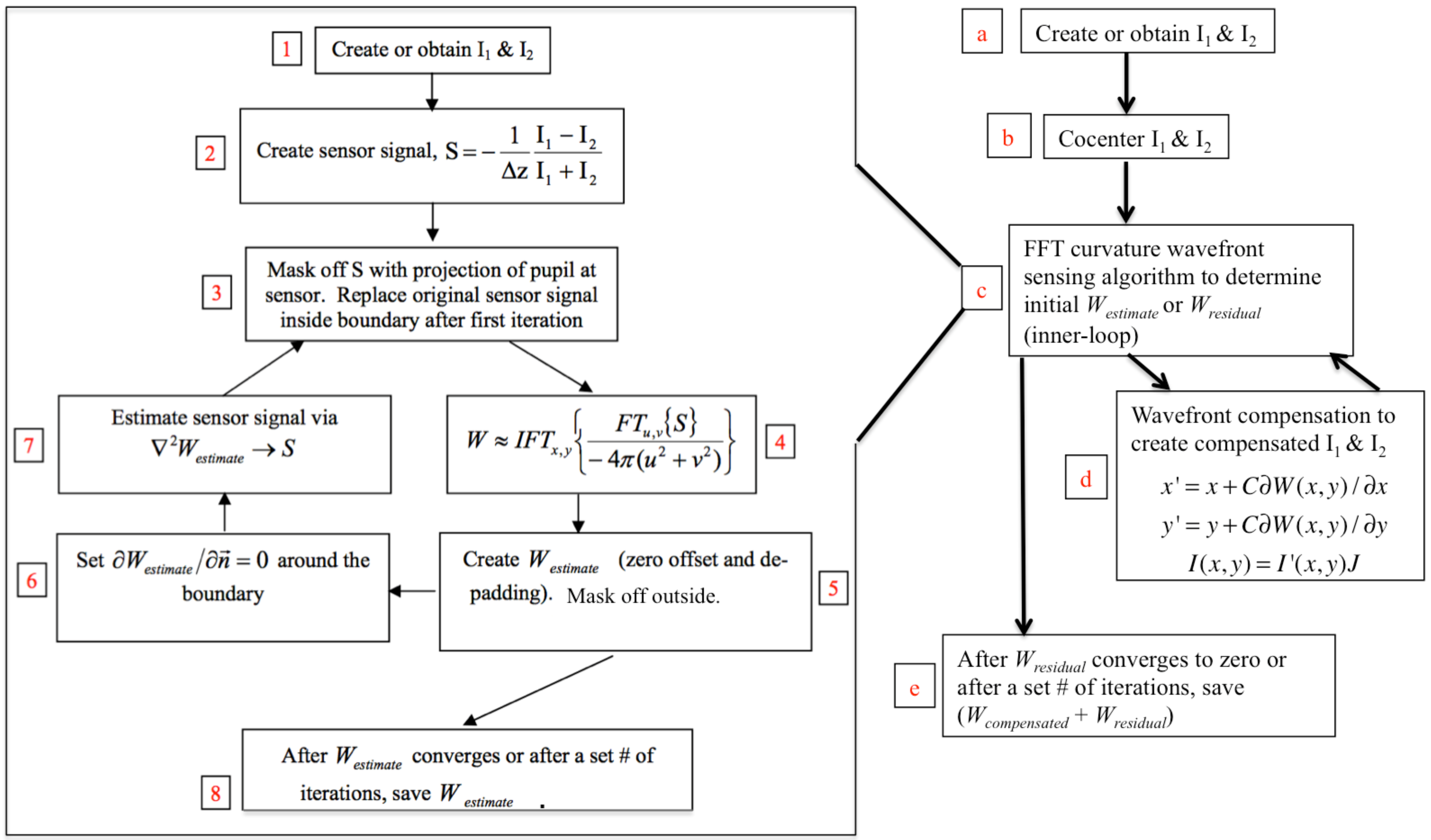}
 \end{tabular}
  \end{center}
  \caption
 { \label{fig:wcflow} 
The block diagram of the iterative FFT algorithm including the ``outer
loop'' image compensation for the 
estimated wavefront aberrations.
}
  \end{figure*}

The implementation of the FFT algorithm is shown in
Figure~\ref{fig:wcflow} (left, boxes 1--8).  
It involves iteratively applying Equation~\ref{eq:wcs} to estimate the
wavefront $W$, then setting boundary condition and putting the
original wavefront signal, $S$, back inside the boundary to obtain a new
signal.  The algorithm we have described thus far is called the ``inner loop".

As was noted by Roddier and Roddier~\cite{Roddier93},
the method above is only a first order approximation valid for
small $\Delta z$ values, i.e., highly defocused images (see Eq.~(\ref{eq:deltaz})).
The initial solution of the Poisson equation we obtain by the ``inner
loop" should be used as
a first order solution that is further refined in a second iterative process.

The overall algorithm accuracy can be improved by iteratively removing or
compensating for the
effects of the estimated wavefront aberrations on the original intra-
and extra-focal intensity images
and then reapplying the ``inner loop" to the corrected images.  This
``outer loop" (Figure~\ref{fig:wcflow} (right, boxes a--e)) is
iterated on until the noise level or a given number of iterations is
reached.  
With each iteration of the
``outer loop" the estimated residual wavefront is summed with the
previous solution to provide the current wavefront estimate.

We compensate/correct the intra- and extra-focal images for
the aberrations of the current wavefront estimate by remapping the image
flux using the Jacobian of the wavefront in the pupil plane.  For this
process we let $R$ be the pupil radius of the
telescope and denote the reduced coordinates in the pupil plane as $x=U/R$ and
$y=V/R$, 
where $U$ and $V$ are Cartesian coordinates in the pupil plane, 
and those in the image plane as $x'$ and $y'$. 
The reduced coordinates in the pupil and image planes are related by~\cite{Roddier93}
\begin{eqnarray}
x' & = & x + C \partial W(x,y)/\partial x, \label{eq:xyp2x}\\
y' & = & y + C \partial W(x,y)/\partial y, \label{eq:xyp2y}
\end{eqnarray}
with
\begin{equation}
C= - \frac{f(f-l)}{l} \frac{1}{R^2}.
\label{eq:C}
\end{equation}
The intensities in the two planes are related by the Jacobian ($J$) due to flux
conservation,

\begin{equation}
I(x,y)/I'(x,y)=J=
\left|
\begin{array}{cc}
\partial x'/ \partial x & \partial x'/ \partial y \\
\partial y'/ \partial x & \partial y'/ \partial y 
\end{array}
\right|
\label{eq:jacobian}
\end{equation}

Using these relations, given a wavefront estimation from a pair of
defocused images, we are able to ``restore'' the pair of images to a
state where the given estimated wavefront aberrations are absent.
The new images are then fed to the inner loop (see box c in
Figure~\ref{fig:wcflow}) where the residual wavefront error is estimated.
The loop continues until the residual wavefront $W_{\rm residual}$ is
consistent with zero, or a given number of outer iterations is reached.

Our compensation algorithm has an oversampling parameter, which enables
sub-pixel resolution for the mapping between the pupil and image planes.
We can choose to improve the compensation performance by increasing
the sub-pixel sampling, but at a cost of computation time.
Due to the fact that the compensation is based on geometrical
ray-tracing, we always compensate on the original defocused images.
A feedback gain less than unity is used to prevent large oscillation
in the final wavefront estimation, i.e., upon each outer iteration
only part of the residual is compensated.
To decouple Zernike terms with the same azimuthal frequency, for
example, between tip-tilt and astigmatisms, and defocus and spherical
aberration, only a certain number of low order Zernike terms are
compensated at each outer iteration. 
After the algorithm has converged on
low order terms, higher order terms are added to the compensated wavefront.
The tests we show in Section~\ref{sec:tests} each include 14 outer
iterations, with the highest Zernike index of 4, 4, 6, 6, 13, 13, 13,
13, 22, 22, 22, 22, 22, and 22, respectively.

The compensation algorithm also helps identify cases where the
geometric limit has been reached. When the compensation procedure
results in a negative intensity on the pupil plane, an in-caustic
warning flag is set. 
Here ``in-caustic'' means rays approaching focus from different points
in the pupil cross before the intra focal or beyond the extra focal
image plane. This crossing leads to ambiguities in the interpretation
of the image intensities and can lead to non-physical negative
intensities in the compensated image. 

\subsection{Series Expansion}
The series expansion method of solving the TIE is based on the
decomposition of the TIE into a series of orthonormal and complete
basis functions~\cite{GuNu96}. Since the outer loop is using the
annular Zernike polynomials, it is natural to choose those for the basis, albeit
the method would certainly work with any orthonormal basis set. 

Let $\Omega$ be the area of the plane with positive intensity $I$ and
smooth boundary $\Gamma$.
As such, the TIE can also be written as:
\begin{eqnarray}
\partial_z I &= & -\nabla\cdot(I\nabla W) \label{eq:tie2}\\
I(x,y) & > & 0 \;\;\; {\rm inside}\;\Omega \\
I(x,y) & = & 0 \;\;\; {\rm outside}\;\Omega\;{\rm and}\; {\rm on}\;\Gamma.
\end{eqnarray}
Let $Z_i(x,y)$, $i$=1,2,3... be a set of orthonormal and complete basis functions
over the pupil,
\begin{equation}
W(x,y)=\sum_{i=1}^{\infty}W_i Z_i(x,y).
\end{equation}
Now we multiply Eq.~(\ref{eq:tie2}) by $Z_j(x,y)$ and integrate over
$\Omega$,
\begin{equation}
\int\int (\partial_z I) Z_j d\Omega = -\int\int \nabla\cdot(I\nabla W) Z_j d\Omega.
\end{equation}
Integrating by parts, and by taking into account the boundary condition,
we get
\begin{equation}
\sum_{i=1}^{\infty} M_{ji} W_i = F_j,
\end{equation}
where
\begin{eqnarray}
F_j&=&\int\int (\partial_z I) Z_j d\Omega\\
M_{ji}&=&\int\int I(x,y) \nabla Z_j \cdot \nabla Z_i d\Omega.
\end{eqnarray}
Therefore,
\begin{equation}
{\mathbf W} = {\mathbf M}^{-1} {\mathbf F}.
\end{equation}

Each of our baseline algorithms have certain advantages and
applications over the other.  
The wavefront compensation we have discussed for the
iterative FFT method can also be used to improve the performance of
the series expansion technique, or any other algorithm used to solve the
TIE.
For detailed analysis of the wavefront
structure the iterative FFT is preferred since it results in a 2-D
image of the wavefront.  
Both methods work with arbitrary pupil geometry, provided that a set
of orthonormal basis functions over the pupil can be found. 
When such basis functions are not available, the accuracy of the series expansion method
degrades more because the expansion relies on the orthogonality of the
basis, whereas for iterative FFT, the non-orthogonality only makes
decomposition of the solved wavefront into the Zernike space problematic.
On the other hand, when speed is a concern, the
series  expansion is the preferred method.  
With the same number of outer iterations, 
we found that
the series expansion is
about 5 times faster than the iterative FFT.
We anticipate both methods
will be used in LSST during commissioning and engineering (iterative
FFT) and routine survey operations (series expansion).
In the analyses presented in the rest of the paper, we use the
series expansion as the inner-loop, and the wavefront compensation as
the outer-loop.

\section{Algorithm Modifications for LSST}
\label{sec:extension}

LSST's optical system poses four algorithmic challenges to using curvature
sensing and solving the intensity transport equation for estimating
the wavefront error at the entrance pupil: 1) the high central
obscuration requires the use of annular Zernike polynomials as the basis
set; 2) the fast $f$/1.23 optical beam results in significant
nonlinearity in projecting the wavefront error on the pupil plane;
3) at the location of the wavefront sensors
there is significant pupil distortion and vignetting that must be
accounted for; and 4) field-dependent variations in the wavefront over
the wavefront sensor area must also be accounted for. Our solutions
for each of these are given in the discussion that follows.

\subsection{Large Central Obscuration}

\begin{figure}[tb]
  \begin{center}
  \begin{tabular}{c}
\includegraphics[width=.8\columnwidth]{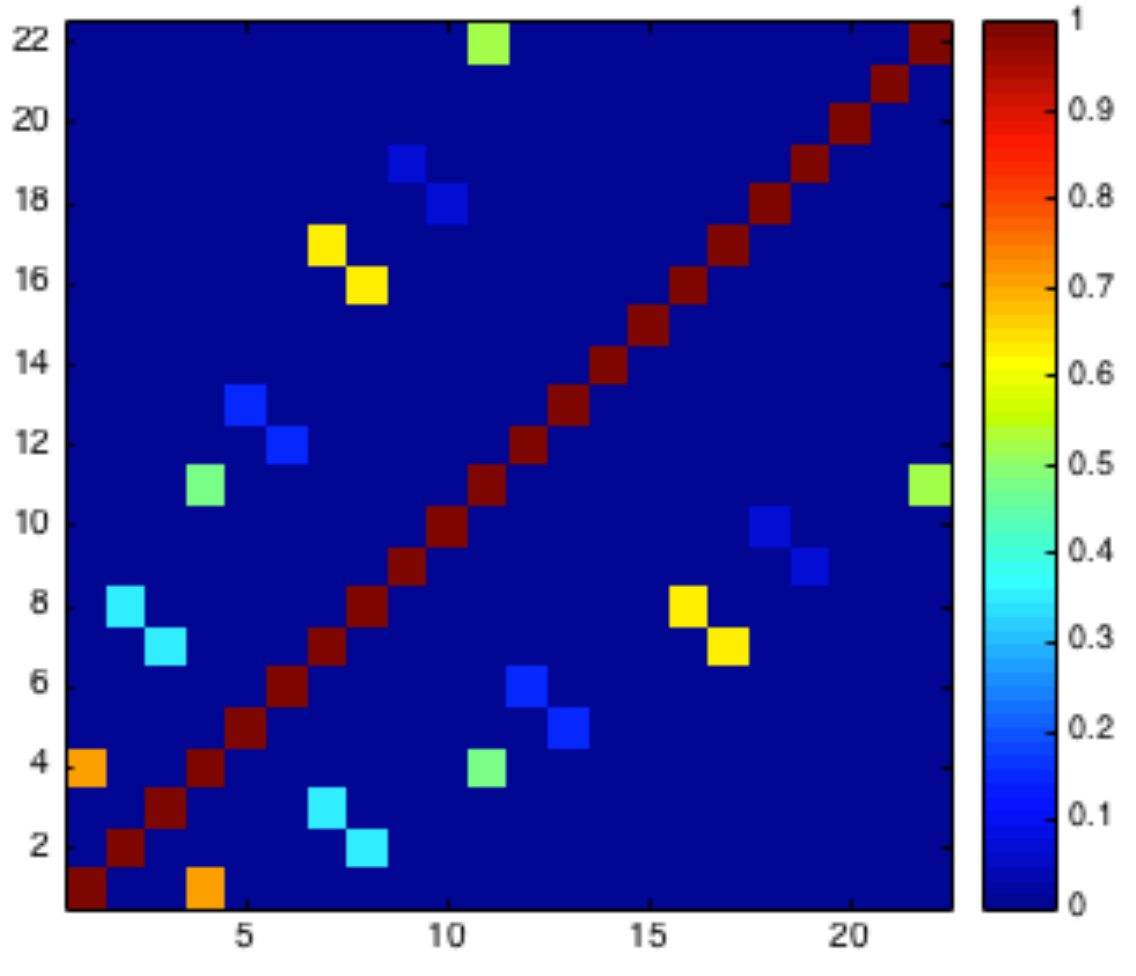}
 \end{tabular}
  \end{center}
  \caption
 { \label{fig:innerp} 
Orthogonality test of the standard Zernike polynomial on an
entrance pupil with 60\% obscuration. The $x$- and $y$- axes are the
order of the Zernike polynomials (up to 22).
}
  \end{figure}

In both the iterative FFT and the series expansion methods, we
decompose the wavefront onto a set of orthonormal and complete basis
functions.
Traditionally, the ``standard'' filled aperture Zernike polynomials
have been chosen for this
purpose due to their well-understood properties~\cite{standardZ}.
When the optical system has non-zero obscuration, the
Zernike polynomials are no longer orthogonal.
When the obscuration is small, the non-orthogonality does not
cause much of a problem. 
However, when the central obscuration gets large, the 
non-orthogonality causes a degeneracy problem in the Zernike
decomposition.  Figure~\ref{fig:innerp} shows the orthogonality test of the standard
Zernike polynomial on an LSST-like 60\% obscured entrance pupil.

The obvious solution is to use the annular Zernike
polynomials~\cite{annularZ}. This ensures the orthogonality on the annular pupil.
However, note that an obscuration means there is a loss of information. 
Even with the annular Zernike polynomials, large central
obscuration means that the wavefront is harder to measure
compared to the case when there is no or small obscuration.

\subsection{Correcting for Fast $f$-number}

In a fast-beam optical system like LSST the wavefront error defined on
the reference sphere at the exit pupil can no longer be projected on
the pupil plane in a straightforward, linear way.  The steepness of
the fast-beam reference sphere results in non-linear mapping from the pupil ($x-y$ plane) to
the intra- and extra-focal images ($x'-y'$ plane).  Consequently, the intensity
distribution on the aberration-free image is no longer uniform and
would give rise to anomalous wavefront aberration estimates.
This effect also causes the obscuration ratio as seen at the intra- and extra-focal image
planes to be different from that on the pupil; thus the linear
equations for the mapping between $x-y$ and $x'-y'$ 
(Eqs.~\ref{eq:xyp2x} and \ref{eq:xyp2y}) are no longer valid.
Therefore, we need to quantify the non-linear fast-beam effect, and be able
to remove it so that the corrected images can be processed using the
iterative FFT or the series expansion algorithms.

For an on-axis fast-beam system, it can be shown that 
\begin{eqnarray}
x'&=&F(x,y) x 
+C \frac{\partial W}{\partial x },\\
y'&=&F(x,y) y 
+C \frac{\partial W}{\partial y }, \label{eq:Aby}
\end{eqnarray}
with
\begin{eqnarray}
F(x,y)=\frac{m\sqrt{f^2-R^2}}{\sqrt{f^2-(x^2+y^2)R^2}},
\end{eqnarray}
where $m = R'f/(lR)$ is the mask scaling factor,
$R'$ is the radius of the no-aberration image, 
$C$ has been defined in Eq.~(\ref{eq:C}),
and
the reduced coordinates
on the image plane are normalized using the paraxial image radius
$R'/m=lR/f$, i.e., the radius of the
image for a paraxial system with the same $f$ and $R$.
Note that flux conservation (Eq.~(\ref{eq:jacobian})) still holds.
The partial derivatives of the Jacobian become
\begin{eqnarray}
\frac{\partial x'}{\partial x} 
& = &
F(x,y)
(1+ \frac{x^2R^2}{f^2-r^2R^2} ) +C
\frac{\partial^2 W}{\partial x^2 } \\
\frac{\partial y'}{\partial y} 
& = &
F(x,y)
(1+ \frac{y^2R^2}{f^2-r^2R^2} ) +C
\frac{\partial^2 W}{\partial y^2 }  \\
\frac{\partial x'}{\partial y} 
& = &
F(x,y)
\frac{xyR^2}{f^2-r^2R^2} +C
\frac{\partial^2 W}{\partial xy } \\
\frac{\partial y'}{\partial x} 
& = &
F(x,y)
\frac{xyR^2}{f^2-r^2R^2} +C
\frac{\partial^2 W}{\partial xy }.
\end{eqnarray}
With this mapping we are able to fully account for geometric
projection effects of the LSST's fast 
$f$-number.

\subsection{Off-axis distortion and Vignetting Correction}

The LSST wavefront sensors are located about 1.7$^\circ$ off the optical axis.
On top of the non-linear projection effects due to the small $f$-number, the
off-axis images are distorted and non-axisymmetric.
Since an analytical mapping from the telescope aperture to the defocused
image planes is not readily achievable, we instead have developed a
numerical solution, representing the mapping between the two sets of
coordinates with 2-dimensional 10th order polynomials. 
The coefficients of the polynomials are determined using least-square
fits to the ray-hit coordinates from a grid of rays simulated by the
LSST ZEMAX model.
The agreement between the fit and the ZEMAX data is better than one
percent of a pixel.
The gradients and Jacobians are then calculated and implemented in the
wavefront compensation code. 

Compared to a paraxial model with the same $f$-number, the off-axis
distortion shifts the ray-hit coordinates on the image planes by up to
about 6 pixels, relative to the chief ray, had there been no central obscuration.
At the upper right corner of the focal plane, this distortion is
symmetric about the 45$^\circ$ line.
In contrast, the maximum shifts in the ray-hit
coordinates on the image plane caused by 200nm of 45$^\circ-$
astigmatism is about 0.2 pixel.

\begin{figure}[tb]
  \begin{center}
  \begin{tabular}{c}
\includegraphics[width=.8\columnwidth]{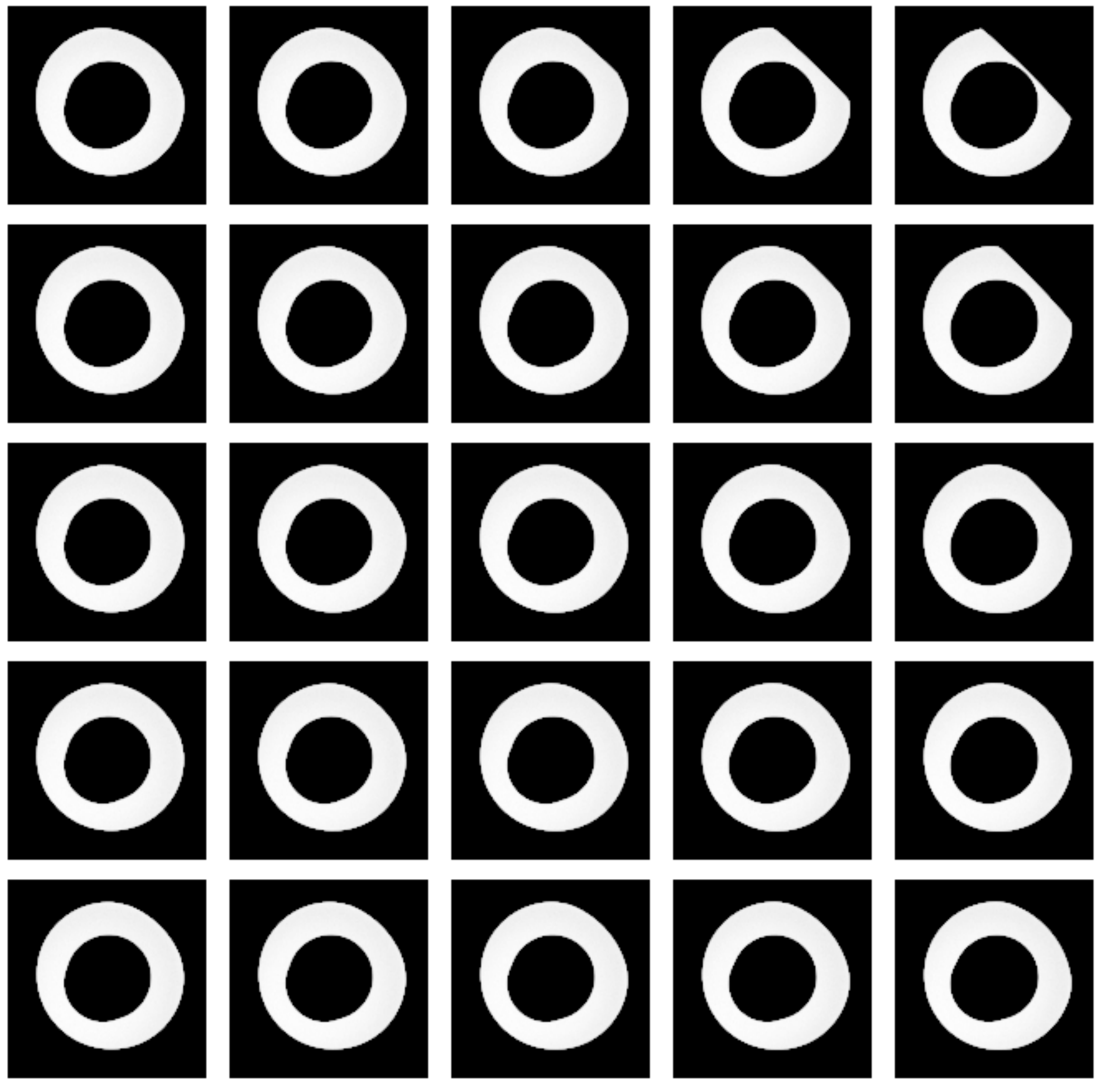}
 \end{tabular}
  \end{center}
  \caption
 { \label{fig:grid_images} 
Intra-focal images on a 5$\times$5
grid covering the LSST upper right wavefront chip with field angle ranging
from 1.51$^\circ$ to 1.84$^\circ$.
Note that these are ZEMAX simulated images for analysis and testing
purposes. In reality, the intra-focal images are acquired on half of
the wavefront chip only.
}
  \end{figure}

Because the wavefront sensors are at the edge of the 3.5$^\circ$ field
of view, the wavefront images are vignetted.
For most parts of the wavefront chip, vignetting is mainly due to
the primary and secondary mirrors and increases gradually with field angle.
When the field angle gets to about 1.7$^\circ$, vignetting due to the
camera body starts to cause a sharp decrease in the fraction of
unvignetted rays.
Figure~\ref{fig:grid_images} shows intra-focal images on a 5$\times$5
grid covering the upper right wavefront chip with field angle ranging
from 1.51$^\circ$ to 1.84$^\circ$.

Vignetting means loss of information on the edge of the pupil,
making it harder to recover the wavefront. 
This is especially true for aberrations like coma, where most of the
intensity variations occur near the edge of the image.
Furthermore,
although neither the iterative FFT nor the series expansion algorithm
has requirements on the shape of the pupil, they do rely on
decomposition of the wavefront onto the annular Zernike polynomials,
which are no longer orthogonal on vignetted pupils.
Because the vignetting is relatively small, $\sim$10\%, the deterioration of
the wavefront estimation accuracy due to this
non-orthogonality has been observed to be small.
At the center of the wavefront chip, unit tests using ZEMAX images
show no visible increase in estimation uncertainty compared to the
on-axis tests.

\subsection{Field-Dependent Corrections}

The split sensor design of LSST introduces two
additional potential sources for error. 
First, the TIE used in curvature sensing requires that
the total flux in the intra- and extra-focal images are the same. Using
different sources for the intra- and extra-focal images in general will violate
this requirement. This can be largely overcome by local background
subtraction and normalizing to the source flux during the processing of the intra- and
extra-focal images prior to solving for the wavefront. 

Second, the curvature sensing method assumes that both the intra- and extra-focal
images contain the same optical properties (aberrations and pupil
geometry). Because the intra- and extra-focal sources will come from
different parts of the field of view they will not have identical
optical properties.  

The off-axis geometric distortions are field-dependent as are the telescope intrinsic 
aberrations and vignetting.  The off-axis correction we discussed
previously corrects intensity variations, regardless of their source,
be it from the off-axis distortion or the field-dependent telescope
intrinsic aberrations.  When we make the off-axis correction, we always
do so referenced to a common field coordinate at the center of the
wavefront sensing area, regardless of the source's original location.
Thus, the off-axis correction removes both the effects of field
dependent geometric distortion and intrinsic design aberrations.

The vignetting pattern still varies by field.  To avoid anomalous
signal due to the difference in vignetting from
entering the wavefront estimation results, we mask off the 
intra- and extra-focal images using a common pupil mask, which is the
logical AND of the pupil masks at the intra- and extra-focal field positions.

\section{Algorithm Validation and Unit Testing}
\label{sec:tests}

We have applied many levels of unit testing on the LSST wavefront
sensing algorithms using images with and without the effects of
atmospheric turbulence. Examples of these tests and their results are
provided below.

\subsection{Paraxial Unit Testing}

\begin{figure}[tb]
  \begin{center}
  \begin{tabular}{c}
\includegraphics[width=0.9\columnwidth]{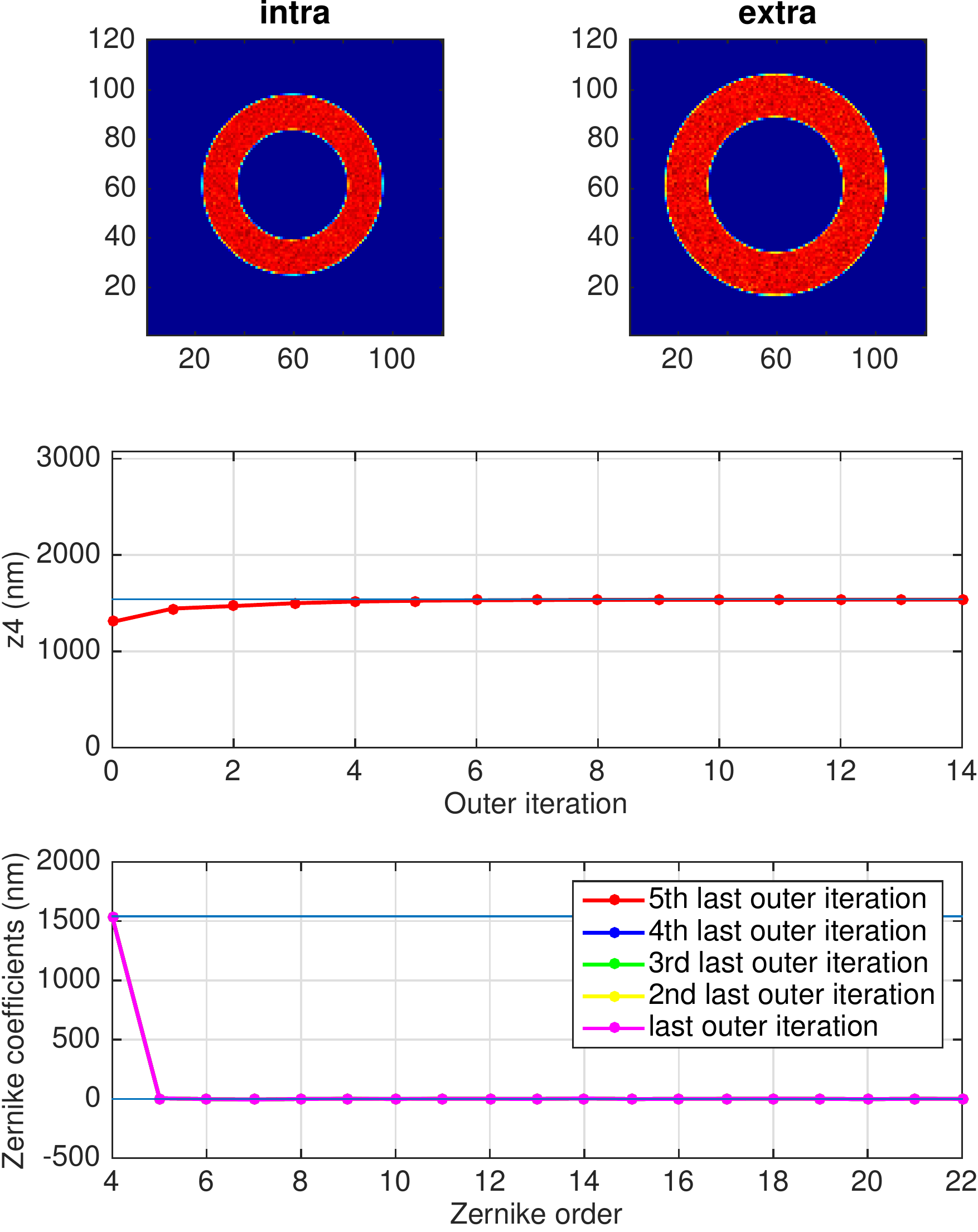} 
\end{tabular}
  \end{center}
  \caption
 { \label{fig:wcsParaxial} 
An example of a paraxial lens unit test for
focus (Z4). The top row shows the input images.
The middle row shows how the estimated Z4 converges to the input
value. The final annular Zernike composition is shown at the bottom.
Note that on the bottom plot, we show results from the last five outer iterations.
These curves are on top of each other, which shows good convergence.
}
  \end{figure}

We have performed comprehensive tests with on-axis images created in ZEMAX
using a paraxial lens model, with $f$-number ranging from 1.3 to 4;
image space defocus from 1mm to 5mm; and obscuration ratios from 0
(filled) to 60\%. 
One example of such tests is given in Figure~\ref{fig:wcsParaxial}.
The results agree nicely with the ground truth from ZEMAX.

Our wavefront estimation algorithm performs very well.
For the single-Zernike-term tests,
each Zernike mode runs into caustic at a different aberration magnitude.
For $f$/1.23, 1mm image space defocus, with RMS wavefront aberration
below 1.5 microns, the only problematic modes are the 5th order astigmatisms.

\subsection{LSST On- and Off-axis}

Unit test results using ZEMAX simulated images 
for various Zernike terms at different magnitudes
show that the fast-beam correction algorithm is very successful. 
To facilitate algorithm testing, the telescope intrinsic aberrations
have been compensated using a phase screen, before applying the known
aberration under test.

\begin{sidewaysfigure*}[tb]
  \begin{center}
  \begin{tabular}{c}
\includegraphics[width=.45\columnwidth]{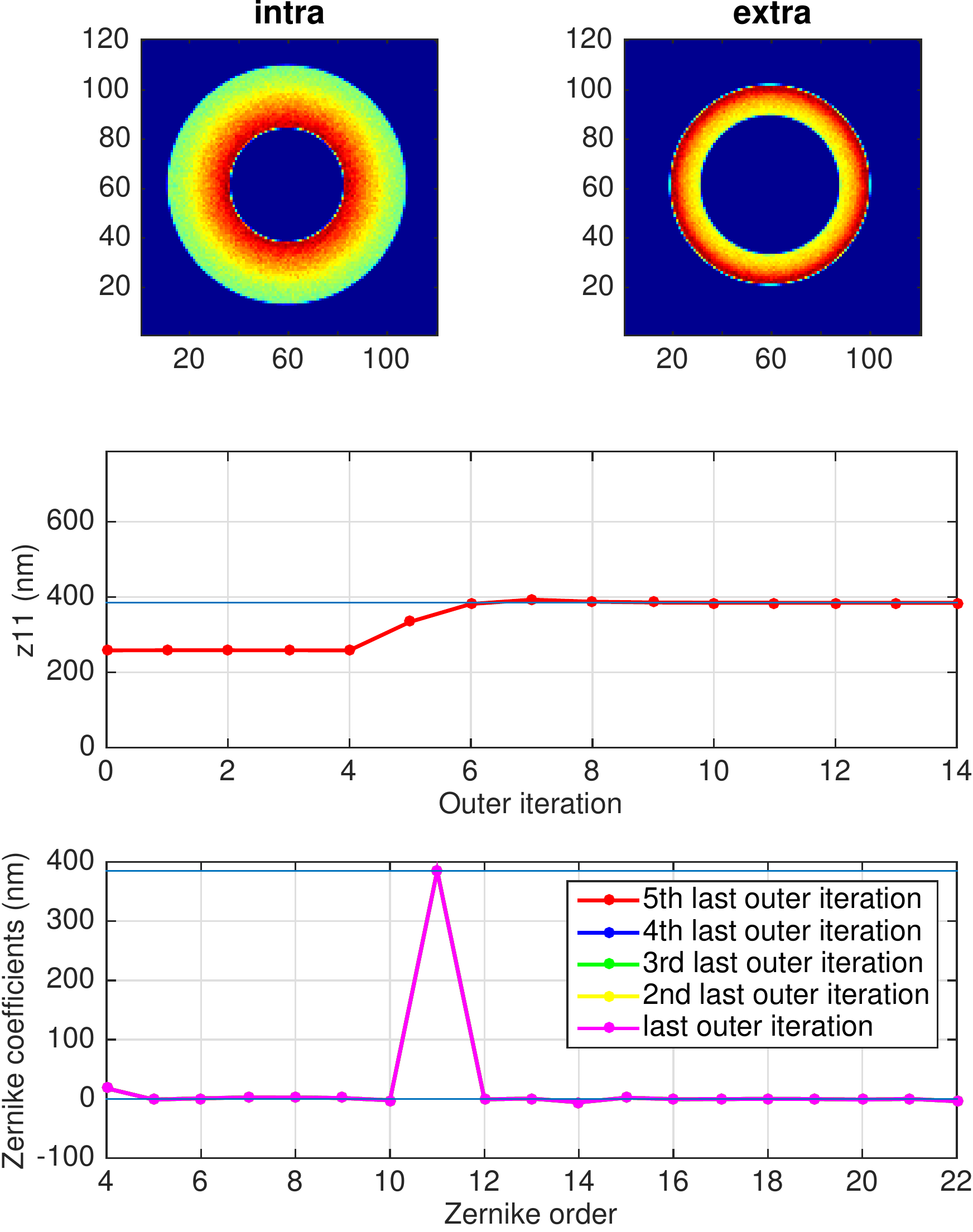} \hspace{5mm}
\includegraphics[width=.45\columnwidth]{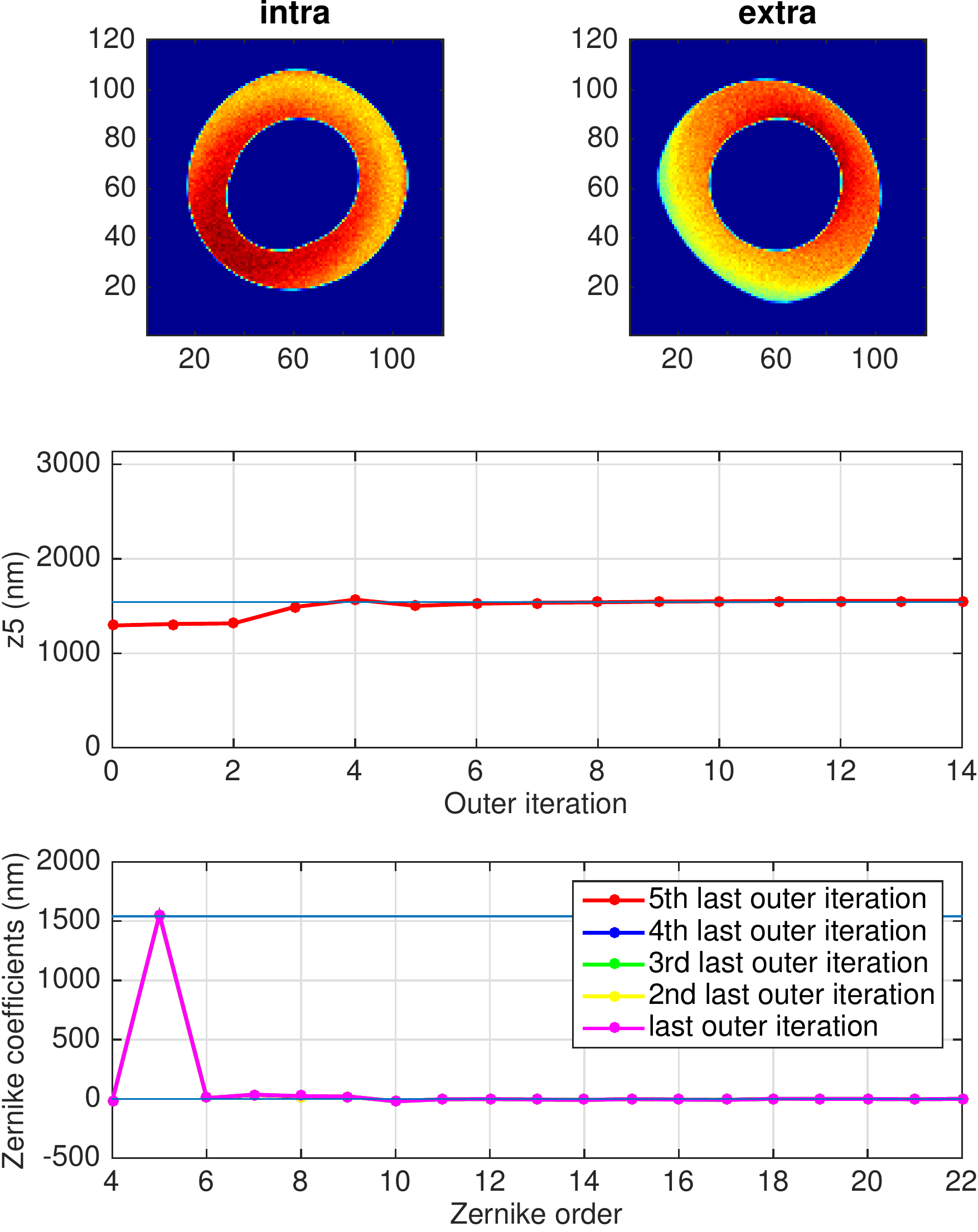} 
\end{tabular}
  \end{center}
  \caption
 { \label{fig:wcsLSST} 
Additional unit testing is done using a ZEMAX model of the LSST
optical system. 
Two examples are shown for spherical aberration (Z11) on-axis (left) 
and astigmatism (Z5) at the center of one of the off-axis wavefront
sensors.
The formatting of the figure is the same as Figure~\ref{fig:wcsParaxial}.
Note that Z11 is compensated in the outer loop starting the 5th
iteration, and Z5 starting the 3rd iteration.
Note that on the bottom plots, we show results from the last five outer iterations.
These curves are on top of each other, which shows good convergence.
}
  \end{sidewaysfigure*} 

One example of such tests is given in Figure~\ref{fig:wcsLSST}.
The intra- and extra-focal images are taken at the center of the focal plane.
In this example, the aberration present is 0.5 wave (wavelength is 770nm) of Z11
(spherical aberration). 
The estimated Z11, as shown in Figure~\ref{fig:wcsLSST}, 
agrees nicely with the ground truth.
All other Zernike terms have magnitudes 
of no more than $\sim$20--30nm.

We have also performed unit tests using ZEMAX simulated LSST off-axis images 
with various Zernike terms at different magnitudes.
One example of such tests,
where 2 waves (wavelength is 770nm) of Z5
(45$^\circ$ astigmatism) is applied, 
is shown in Figure~\ref{fig:wcsLSST}. 
The intra- and extra-focal images are taken at field position
(1.185$^\circ$, 1.185$^\circ$), corresponding to the center of one of
the LSST wavefront sensors. Vignetting is clearly visible in these images.
Again, the input aberration is recovered, as shown in Figure~\ref{fig:wcsLSST}.
All other Zernike terms have magnitudes very close to zero.
Tests are also performed using images simulated by the LSST Photon
Monte Carlo Simulator~\cite{phosim}. A similar level of agreement
between the measured wavefront and the truth is observed.

\begin{figure}[tb]
  \begin{center}
  \begin{tabular}{c}
\includegraphics[width=.9\columnwidth]{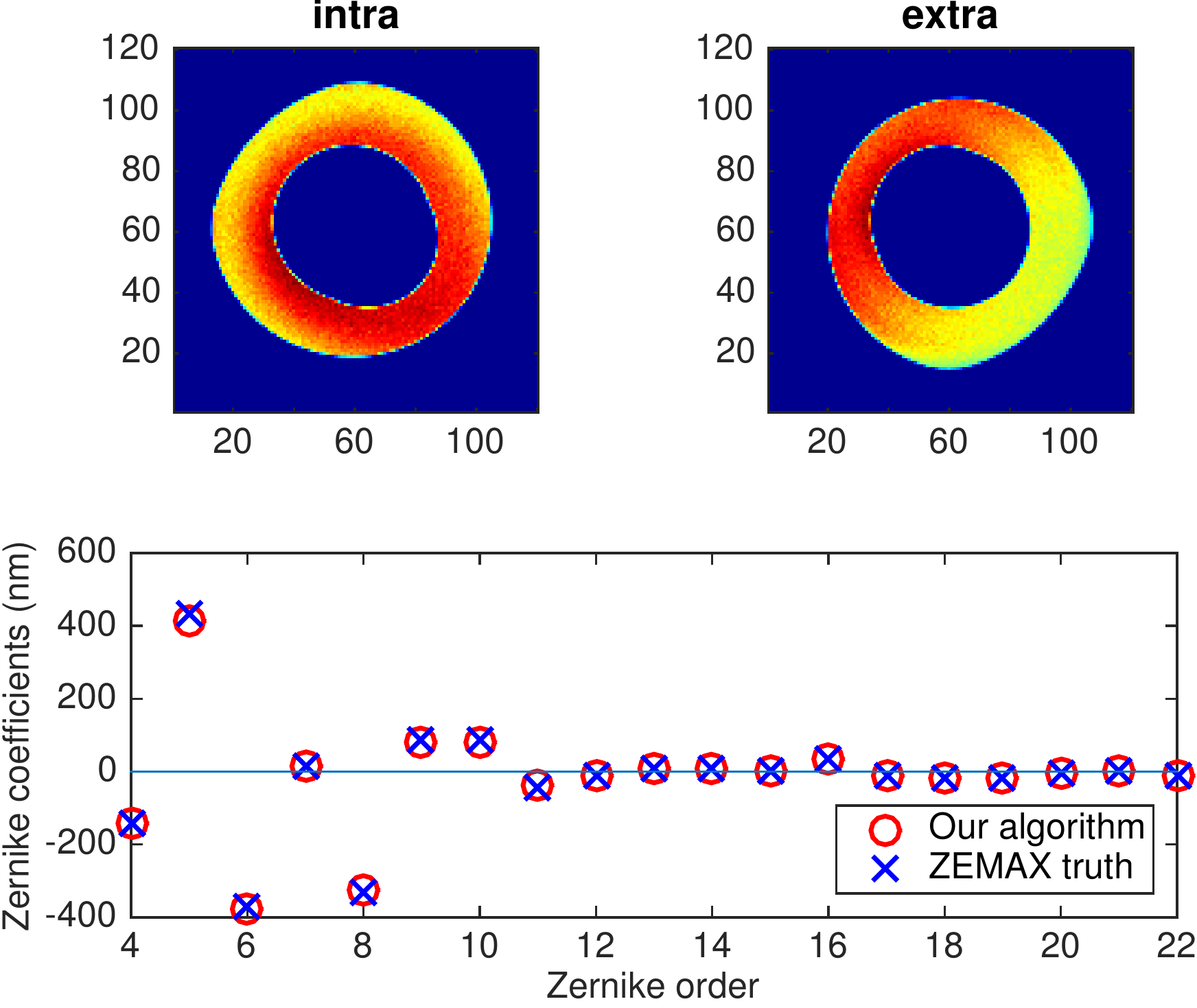} 
\end{tabular}
  \end{center}
  \caption
 { \label{fig:wcsPerturb} 
ZEMAX simulation at the field angle of
($-$1.185$^\circ$, 1.185$^\circ$)
containing all intrinsic aberrations plus a decenter of
0.5mm in the secondary mirror. 
The recovered aberration coefficients (red circles) are in excellent
agreement with the ZEMAX-calculated coefficients (blue crosses). 
}
  \end{figure}

As another set of tests, the ZEMAX model is used 
including all intrinsic aberrations and various individual
perturbations at the 4 corner wavefront sensor locations.  
Figure~\ref{fig:wcsPerturb} shows an example at the field angle of
($-$1.185$^\circ$, 1.185$^\circ$), where the secondary mirror has been
decentered by 0.5mm.

\subsection{Algorithm Linearity}

\begin{figure}[tb]
  \begin{center}
  \begin{tabular}{c}
\includegraphics[width=.9\columnwidth]{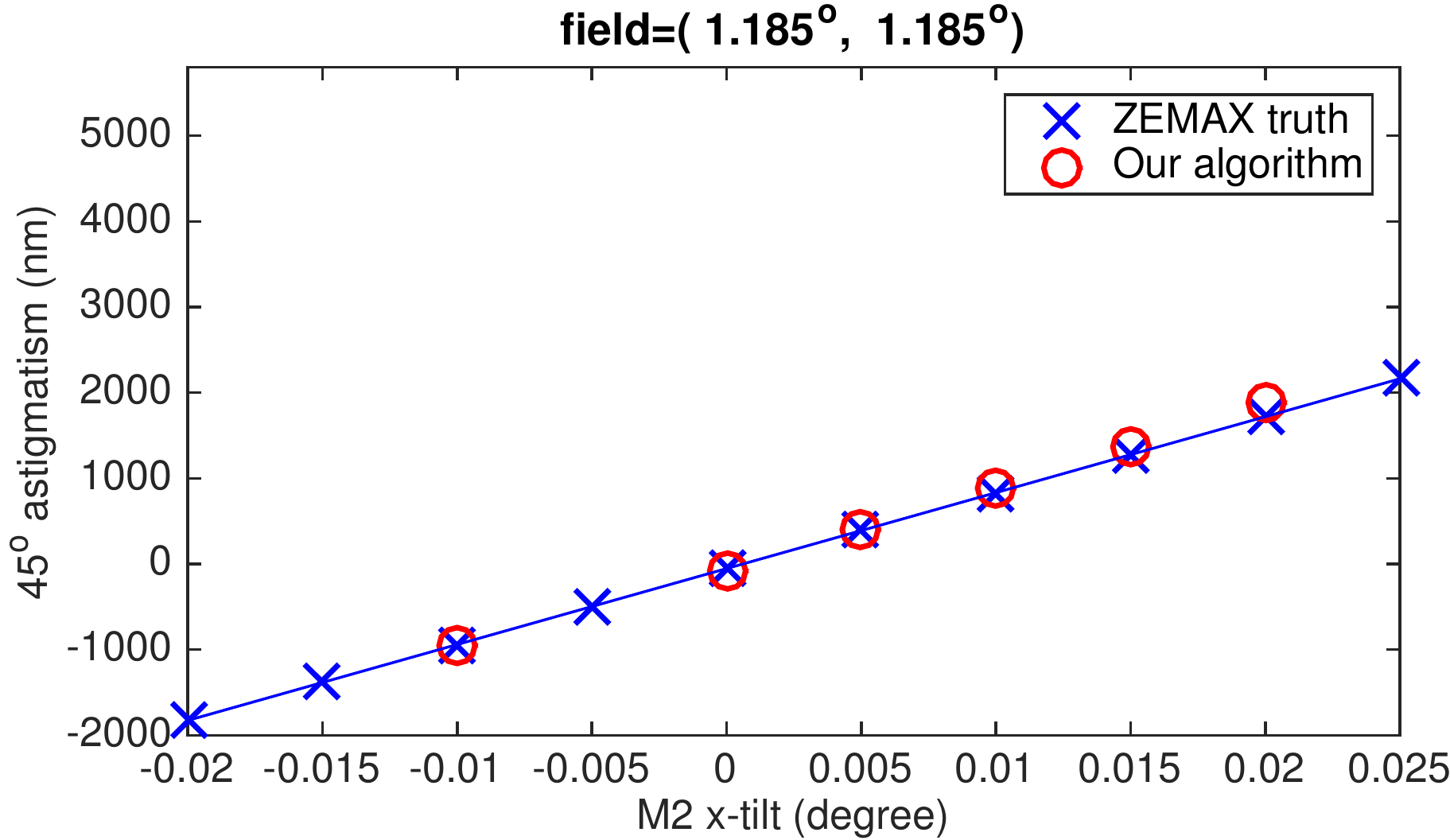} 
\end{tabular}
  \end{center}
  \caption
 { \label{fig:linearity} 
ZEMAX-calculated wavefront (blue crosses) 
and the recovered wavefront (red circles)
at one of the four corners of the focal plane. 
The recovered aberration coefficients start to deviate from
linearity in proximity of the geometric limit.
}
  \end{figure}

By varying the controlled degrees of freedom of the telescope model and repeating
the wavefront measurements, good
algorithmic linearity has been observed within the geometric limit.
One example of such tests is shown in Figure~\ref{fig:linearity}.
When the secondary
mirror is tilted by incremental amounts, and all other degrees of
freedom of the telescope stay unperturbed,
the wavefront Zernikes (45$^\circ$-astigmatism shown) change linearly.
The linearity starts to degrade in proximity of the geometric limit.
Similar behavior has also been observed for other degrees of freedom
of the system.

\subsection{Covariance Analysis}

The atmospheric covariance is computed for 15-second time integrated
atmospheres generated using the Arroyo library~\cite{arroyo}.
Six layers of Kolmogorov phase screens at various heights are
simulated. Based on historic DIMM data at Cerro Pachon, and the
current understanding on its outer scale, we use $r_0=17$cm and assume
infinite outer scale. 

\begin{figure}[tb]
  \begin{center}
  \begin{tabular}{c}
\includegraphics[width=.9\columnwidth]{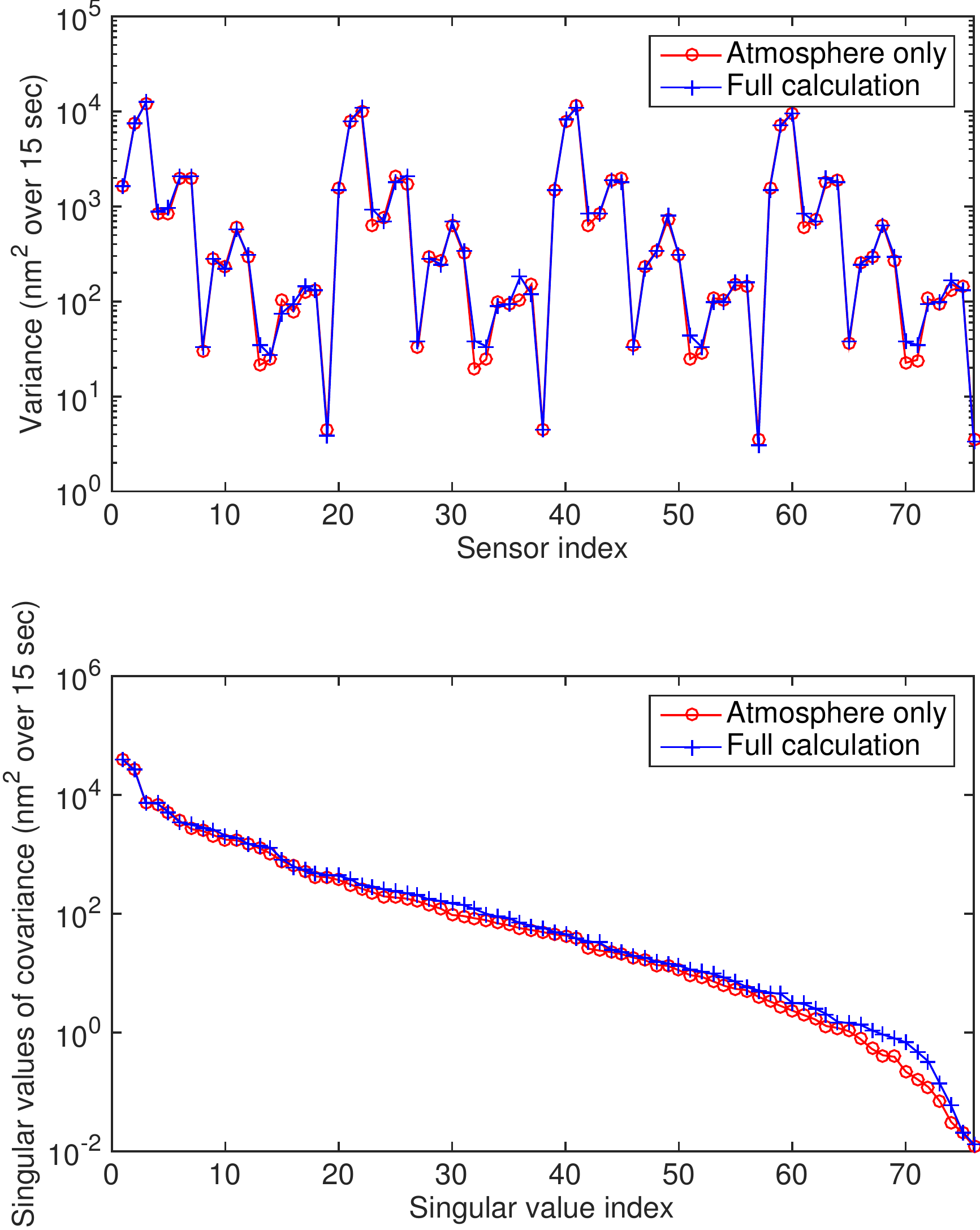} \\
\end{tabular}
  \end{center}
  \caption
 { \label{fig:cov} 
The nearly identical covariance and singular values of the ideal
sensor (atmosphere only) and fully simulated and processed images
show that the noise from the atmosphere is dominant.
The sensor index ranges between 1 and 76, corresponding to the wavefront Zernike coefficients Z4--Z22
at the four corners.
}
  \end{figure}

The same time-integrated phase screens used for the atmosphere
covariance are then applied to a ZEMAX ray trace model in order to
evaluate the importance of algorithm noise with respect to atmosphere
noise in estimating the wavefront. For each time-integrated phase
screen instance, ZEMAX is used to generate intra- and extra-focal
image pairs at each of the 4 wavefront sensors. These images are then
processed through the LSST specific curvature algorithms to estimate
the wavefront coefficients. Compared to the calculated
coefficients from atmosphere alone (ideal sensor),
the results from estimating the wavefront using the simulated sensors
are similar, indicating that the
covariance is dominated by the atmospheric contribution and not the
wavefront sensing algorithms. The plots in Figure~\ref{fig:cov} show the
diagonal entries (19 entries for each field point, corresponding to the Zernike terms estimated)
and the singular values of the covariance matrix; the largest singular values most important
for estimating the misalignment of the telescope are almost identical.

\section{Summary, Conclusions, and Future Work}

Extensions to two curvature wavefront sensing algorithms for LSST have
been developed.  The underlying algorithms considered are the iterative FFT method by the
Roddiers~\cite{Roddier93} and the series expansion based method by Gureyev and Nugent~\cite{GuNu96}.
These are well-established methods that have been proven to work well
for paraxial systems.  Several modifications are needed to make them work for LSST,
to overcome challenges including a highly obscured pupil, the fast 
$f$-number, pupil distortion and vignetting at the field corners, and 
variation of the wavefront over the area covered by the split-sensors.
Our baseline algorithm for use in routine operations is the series
expansion method, due to its higher computational efficiency.

Extensive simulations have been performed using images generated by ZEMAX
and the LSST Photon Monte Carlo simulator~\cite{phosim}.

Integrated modeling of the LSST AOS, including the optimal control of the
system, as well as adapting the algorithms to be used by several other
operating telescopes, is reported in~\cite{AOSspie}.   As the LSST
construction progresses we are also looking into
alternative wavefront sensing algorithms, such as the forward
modeling technique used by DECam, and the PSF-based technique used by VST,
to compare their performance to the results shown here.

This is the first of a series of papers being planned on the Active 
Optics System of LSST. 
The 
following topics are planned to be discussed in subsequent papers 
hence not included here:  
1) the validation of the algorithms presented in this paper using real
data taken at major operating telescopes; 
2) the LSST alignment strategy, the tomographic 
optical reconstruction, and the telescope control algorithm; 3) the 
general functionality of the LSST active optics systems (AOS) and 
the prototype processing pipeline we are using to develop a full 
simulation of the LSST AOS operation. 

\section*{Acknowledgments}

LSST project activities are supported in part by the National Science
Foundation through Cooperative Support Agreement (CSA) Award
No. AST-1227061 under Governing Cooperative Agreement 1258333 managed
by the Association of Universities for Research in Astronomy (AURA),
and the Department of Energy under contract with the SLAC National
Accelerator Laboratory.   Additional LSST funding comes from private
donations, grants to universities, and in-kind support from LSSTC
Institutional Members.

\end{document}